\documentclass[letterpaper,twocolumn,review,showpacs]{revtex4}
\usepackage[latin9]{inputenc}
\usepackage{amsmath}
\usepackage{graphicx}
\usepackage{amssymb}
\usepackage{esint}

\begin{document}

\title{Out of plane screening and dipolar interactions in heterostructures}

\author{Cheung Chan}

\author{T. K. Ng}

\affiliation{Department of Physics, Hong Kong University of Science and Technology,
Clear Water Bay, Kowloon, Hong Kong, China}

\date{\today}
\begin{abstract}
Out-of-plane screening (OPS) is expected to occur generally in metal-semiconductor
interfaces but this aspect has been overlooked in previous studies.
In this paper we study the effect of OPS in electron-hole bilayer
(EHBL) systems. The validity of the dipolar interaction induced by
OPS is justified with a RPA calculation. Effect of OPS in electron-hole
liquid with close-by screening layers is studied. We find that OPS
affects the electronic properties in low density and long wavelength
regime. The corresponding zero-temperature phase diagram is obtained
within a mean field treatment. We argue that our result is in general
relevant to other heterostrucutures. The case of strongly correlated
EHBL is also discussed.
\end{abstract}


\maketitle

\section{introduction}

Modern micro-electronics relies to a large degree on surface science,
which concerns the material properties near a surface or interface.
To enhance the performance of such devices, knowledge of the electronic
states near the interfaces is required. Near a surface or interface,
electronic reconstruction may alter three key factors - interaction
strengths, bandwidths and electron densities \cite{AMillis} which
determine electronic states and their properties.

In this paper, we consider another factor - the modification in form
of interaction between electrons. For instance, in an insulator-semiconductor-insulator
superstructure, if the dielectric constant of the semiconductor is
sizably larger than that of insulator (barrier layer), the image charges
induced at semiconductor-insulator interface can substantially enhance
the binding energy of the excitons confined in the semiconductor layer
\cite{Xconfine95,Xconfine92}. In this case, the electrons and holes
do not interact via usual Coulomb potential after the effect of the
image charges at the semiconductor-insulator interface is taken into
account.

Recently, Huang \textit{et al.}\ observed non-activated electronic
conductivity of a two-dimensional (2D) low density hole system in
a heterojunction insulated-gate field-effect transistor \cite{Nonact-transport}.
Such non-activated conductivity is unexpected as at low charge density
strong Coulomb interaction is expected to crystallize the system (Wigner
crystal), which is then pinned by disorder resulting in insulating
behavior and activated conductivity. Huang \textit{et al.}\ attribute
the behavior to the screening of Coulomb interactions by the metallic
gate, which leads to destruction of the Wigner crystal phase. Physically,
the metallic gate which is located at a distance away from the 2D
hole gas, provides an out of plane screening (OPS) to the hole-hole
interaction, resulting in effective dipolar interaction between holes.
Microscopically, when a charge is placed near a metal surface, an
image charge of opposite sign will be induced at the surface to screen
out the (static) electric field from the charge. From elementary electrostatics,
the system can be described equivalently as a dipole formed by the
charge and its image charge and the interaction between two charges
located near the interface changes from a Coulomb potential $\sim1/r$
to a dipolar potential $\sim1/r^{3}$. This modified interaction,
which is generally expected to exist in metal-semiconductor heterostructures,
can change the electronic properties near the interface. Unexpectedly,
there has been no detailed theoretical study of this effect on electronic
properties until recently \cite{HoLH}. The neglect of OPS might be
due to dynamical screening of in-plane charges \cite{HoLH}. For high
charge density, the screening can effectively reduce both Coulomb
and dipolar interactions to short range interactions. However for
low charge density electronic liquids in-plane screening is less effective
and OPS can lead to a difference, as is observed by Huang \textit{et
al.}\ \cite{Nonact-transport}.

In this paper, we study how OPS affects the electronic properties
in systems with two-layer of charges of opposite sign, i.e.\ the
2D electron-hole bilayer (EHBL) system. We shall study how OPS affects
Wigner crystalization and exciton condensate in the system \cite{Nonact-transport,exciton-cond}
and will also comment on the effect of OPS in interfaces between metals
and strongly correlated electron systems \cite{thin_film_on_metal,YBCO_metal_interface1,YBCO_metal_interface2}.

\section{OPS and effective interaction between charges}

\begin{figure}
\includegraphics[width=0.9\columnwidth]{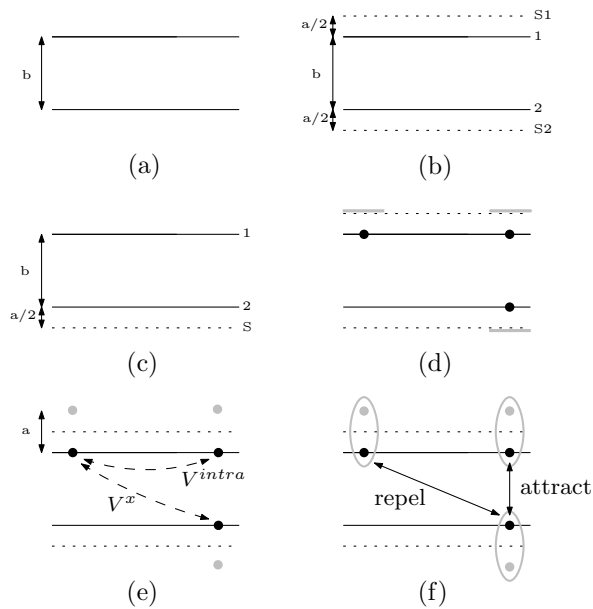}

\caption{\label{fig:EHBL-OPS} (a) EHBL system separated by distance $b$.
(b) EHBL with OPS by metallic plates in both layers. Dotted line represents
metallic interface, separated from the main layer by a distance of
$a/2$. (c) Similar to (b) but with only one OPS layer. (d) Charges
(black dots) and screening charge response at the metal interface
(grey patches). (e) Effective image charge (grey dots) and effective
interactions $V^{\text{intra}}$ and $V^{x}$. (f) Charges attract
when they are aligned while repel when they are not. This behavior
is different from the Coulomb potential which is always attractive
for a pair of electron and hole. The repulsive behavior inhibits exciton
pairing.}

\end{figure}

In this section we provide the details for the EHBL systems we study
and the corresponding OPS effective interaction. We shall assume that
the only effect of the metallic screening layers is to provide an
image charge for point charges sitting close to it and the effective
interaction between charges will be derived from the image charge
picture. The validity of this approximation is bounded by the plasma
frequency $\omega_{p}^{(s)}$ of the screening layer, above which
the screening layer cannot respond rapidly to the charge fluctuations.
Thus our approximation is valid when the plasma frequency of the EHBL
layer $\omega_{p}$ is much less than $\omega_{p}^{(s)}$, or that
the screening layer has density of electric charge much larger than
the charge density of the EHBL layers we consider. The image charge
picture can be justified by a Random Phase Approximation (RPA) calculation
which is shown in the Appendix.

Starting with a EHBL system (Fig\@.\,\ref{fig:EHBL-OPS}(a)), two
metallic screening layers can be added as shown in Fig\@.\,\ref{fig:EHBL-OPS}(b),
or a single metallic screening layer can be added as shown in Fig\@.\,\ref{fig:EHBL-OPS}(c).
We first consider the two-layer case (b). Fig\@.\,\ref{fig:EHBL-OPS}(d)
depicts the charge response in the metallic layer to a nearby charge.
The charge response is assumed to be an image charge, which carries
opposite charge of the same magnitude and is centered at distance
$a$ from the point charge. Thus the point charge and the screening
charge together form a dipole. We have assumed that the distance between
the two layers of charges $b$ is sufficiently larger than $a$ ($b\gg a$)
such that the presence of the other screening layer does not affect
the simple dipole picture. In this case, the intralayer interaction
between two charges located in an OPS layer (Fig\@.\,\ref{fig:EHBL-OPS}(e))
is in real space \begin{equation}
V^{\mathrm{intra}}(\vec{r})=\frac{e^{2}}{\epsilon_{e,h}}\left(\frac{1}{r}-\frac{1}{\sqrt{r^{2}+a^{2}}}\right)\;,\end{equation}
 where $r$ is the charge-charge distance within the charge plane.

It is easy to see that for $r\gg a$, $V^{\mathrm{intra}}$ scales
as $1/r^{3}$ while for $r\ll a$ it follows the usual Coulomb scaling
$1/r$. By using 2D Fourier transform $\frac{1}{\sqrt{r^{2}+a^{2}}}\overset{\mathrm{2D}\mathcal{F}}{\longrightarrow}\frac{2\pi}{k}e^{-ka}$,
the Fourier transformed interaction is \begin{equation}
V^{\mathrm{intra}}(\vec{k})=\frac{2\pi e^{2}}{\epsilon_{e,h}k}\left(1-e^{-ka}\right)\;.\label{v2}\end{equation}
 For an electron and a hole sitting in different layers, the interlayer
interaction is \begin{eqnarray*}
V^{x,2}(\vec{r}) & = & -\frac{e^{2}}{\epsilon_{x}}\left(\frac{1}{\sqrt{r^{2}+b^{2}}}\right.\\
 &  & \left.-\frac{2}{\sqrt{r^{2}+(a+b)^{2}}}+\frac{1}{\sqrt{r^{2}+(2a+b)^{2}}}\right)\end{eqnarray*}
 and its Fourier counterpart is \begin{equation}
V^{x,2}(\vec{k})=-\frac{2\pi e^{2}}{\epsilon_{x}k}e^{-kb}(1-e^{-ka})^{2}\;.\label{v3}\end{equation}
 $\epsilon_{e,h}$ and $\epsilon_{x}$ are the intra-layer and inter-layer
dielectric constants, respectively.

Next we consider EHBL with only one metallic screening layer (see
Fig.\,\ref{fig:EHBL-OPS}(c)). In this case the two layers of charges
have distance $a/2+b$ (layer $1$) and $a/2$ (layer $2$) from the
screening layer, respectively. The intralayer interactions are thus
\begin{eqnarray*}
V_{1}^{\mathrm{intra}}(\vec{r}) & = & \frac{e^{2}}{\epsilon_{1}}\left(\frac{1}{r}-\frac{1}{\sqrt{r^{2}+(a+2b)^{2}}}\right)\;,\\
V_{2}^{\mathrm{intra}}(\vec{r}) & = & \frac{e^{2}}{\epsilon_{2}}\left(\frac{1}{r}-\frac{1}{\sqrt{r^{2}+a^{2}}}\right)\;.\end{eqnarray*}
 with corresponding Fourier transforms \begin{eqnarray}
V_{1}^{\mathrm{intra}}(\vec{k}) & = & \frac{2\pi e^{2}}{\epsilon_{1}k}\left(1-e^{-k(2b+a)}\right)\;,\label{v4}\\
V_{2}^{\mathrm{intra}}(\vec{k}) & = & \frac{2\pi e^{2}}{\epsilon_{2}k}\left(1-e^{-ka}\right)\;.\nonumber \end{eqnarray}
 The corresponding intralayer interaction is given by \[
V^{x,1}(\vec{r})=-\frac{e^{2}}{\epsilon_{x}}\left(\frac{1}{\sqrt{r^{2}+b^{2}}}-\frac{1}{\sqrt{r^{2}+\left(a+b\right)^{2}}}\right)\]
 and \begin{equation}
V^{x,1}(\vec{k})=-\frac{2\pi e^{2}}{\epsilon_{x}k}e^{-kb}\left(1-e^{-ka}\right)\;.\label{v5}\end{equation}

\section{Collective density responses}

In this section we study the collective density responses of the EHBL
systems we considered. For a two component electronic system, the
density-density response of the system is described by a $2\times2$
matrix $\chi_{ij}(q,\omega)$ with $i,j=1,2$. The density-density
response matrix is given in RPA by \cite{G&V}

\begin{equation}
\left(\begin{array}{cc}
\chi_{11} & \chi_{12}\\
\chi_{21} & \chi_{22}\end{array}\right)=\frac{1}{\kappa}\left(\begin{array}{cc}
(1-\chi_{02}V_{22})\chi_{01} & \chi_{01}V_{12}\chi_{02}\\
\chi_{02}V_{21}\chi_{01} & (1-\chi_{01}V_{11})\chi_{02}\end{array}\right)\label{dmatrix}\end{equation}
 where \begin{eqnarray}
\kappa(q,\omega) & = & (1-\chi_{01}(q,\omega)V_{11}(q))(1-\chi_{02}(q,\omega)V_{22}(q))\nonumber \\
 &  & -\chi_{01}(q,\omega)V_{12}(q)\chi_{02}(q,\omega)V_{21}(q)\;,\label{pole}\end{eqnarray}
 $V_{ij}(q)$ is the {}``bare'' interaction between $i^{th}$ and
$j^{th}$ components of the electronic liquid and \begin{equation}
\chi_{0i}(q,\omega)=g_{s}\int\frac{d^{2}k}{(2\pi)^{2}}\frac{n_{F}\left(\varepsilon_{k}^{(i)}\right)-n_{F}\left(\varepsilon_{k+q}^{(i)}\right)}{\hbar\omega+\varepsilon_{k}^{(i)}-\varepsilon_{k+q}^{(i)}}\;,\label{eq:responsefunc}\end{equation}
 where $\varepsilon_{k}^{(i)}\sim k^{2}/2m^{(i)}$ is kinetic energy
of species $i$ particles (of mass $m^{(i)}$), $n_{F}$ is the Fermi-Dirac
distribution function and $g_{s}=2$ is spin degeneracy. In the case
of two screening layers the interactions $V_{11(22)}$ and $V_{12}=V_{21}$
are given by $V^{\text{intra}}(q)$ (eq.\,\eqref{v2}) and $V^{x,2}(q)$
(eq.\,\eqref{v3}), respectively whereas they are given by $V_{1(2)}^{\mathrm{intra}}(q)$
(eq.\,\eqref{v4}) and $V^{x,1}(q)$ (eq.\,\eqref{v5}), respectively
if there is only one screening layer.

Next we study the collective excitations (i.e.\ plasmons) in the
system. The dispersion of the collective excitations are given by
the equation \begin{equation}
\kappa(q,\omega(q))=0\;.\label{eq:plasmoneq}\end{equation}
 We shall first consider the long wavelength limit ($q\rightarrow0$)
where the equation can be studied analytically. In this limit it is
easy to show that \begin{equation}
\chi_{0}\left(q,\omega\right)=\frac{n}{m}\left(\frac{q}{\omega}\right)^{2}+\mathcal{O}\left(\left(\frac{q}{\omega}\right)^{4}\right)\;,\end{equation}
 where $n=\frac{g_{s}(\pi k_{F}^{2})}{(2\pi)^{2}}$ is carrier density.
We have neglected the component index $i$ for brevity.

We begin with the Coulomb case (no screening layer). The interactions
are respectively $V_{1,2}(q)=\frac{2\pi e^{2}}{\epsilon_{1,2}q}$
and $V_{x}(q)=-\frac{2\pi e^{2}}{\epsilon_{x}q}e^{-qb}\sim-\frac{2\pi e^{2}}{\epsilon_{x}q}$
for $q\ll b^{-1}$. The plasmon equation in $q\rightarrow0$ limit
reads \begin{eqnarray}
1-2\pi e^{2}\left(\frac{n_{1}}{m_{1}\epsilon_{1}}+\frac{n_{2}}{m_{2}\epsilon_{2}}\right)\frac{q}{\omega^{2}}+\left(2\pi e^{2}\right)^{2}\nonumber \\
\times\frac{n_{1}n_{2}}{m_{1}m_{2}}\left[\frac{1}{\epsilon_{1}\epsilon_{2}}-\frac{1}{\epsilon_{x}^{2}}\right]\left(\frac{q}{\omega^{2}}\right)^{2} & = & 0\;.\label{eq:Coul plasmon}\end{eqnarray}
 We first consider the case $\epsilon_{x}^{2}=\epsilon_{1}\epsilon_{2}$
such that the term in the square bracket is zero. In this case we
need to expand the interlayer interaction to one order higher in $q$.
As a result the last term in eq.\,\eqref{eq:Coul plasmon} is replaced
by a term of order $\frac{q^{3}}{\omega^{4}}$ and the plasmon equation
at long wavelength limit yields two solutions, which are the out-of-phase
mode ($\omega\sim q$) and in-phase mode ($\omega\sim\sqrt{q}$).
Indeed this occurs usually in a 2D electronic systems with both conduction
and valence bands where the same dielectric constant $\epsilon_{x}^{2}=\epsilon_{1}\epsilon_{2}$
is found for all interactions. In the more general case $\epsilon_{x}^{2}\neq\epsilon_{1}\epsilon_{2}$,
which arises quite naturally in the complex environment of EHBL heterostructures,
we can easily see from eq.\,\eqref{eq:Coul plasmon} that the plasmon
frequency scales as $\omega\sim\sqrt{q}$. There are two modes of
plasmons.

\begin{figure}
\noindent \includegraphics[width=1\columnwidth]{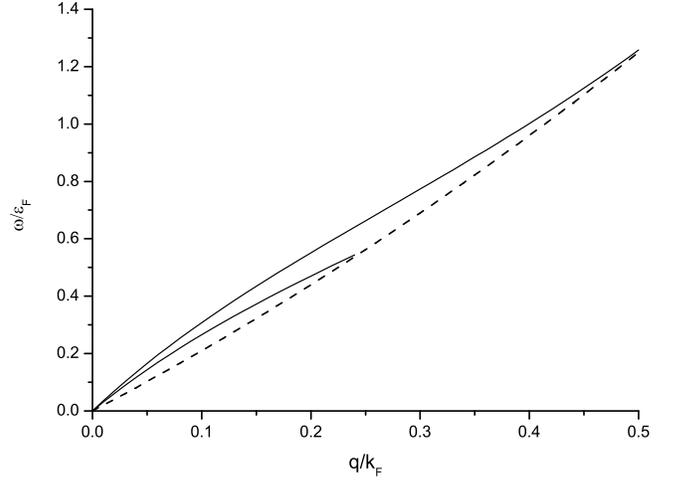}

\caption{\label{fig:Plasmon-d-d}Plasmon excitations in EHBL with two screening
layers computed numerically. The spectrum is computed by solving eq.\,\eqref{eq:plasmoneq}
numerically with the full expression of $\chi_{0}(q,\omega)$ \cite{G&V}.
The solid lines are the plasmons excitations and the dash line represents
the boundary of the particle-hole continuum. The wave-number $q$
and frequency $\omega$ are normalized with respect to Fermi momentum
$k_{F}$ and Fermi energy $\epsilon_{F}$, respectively. We set $a=1$,
$b=15$, $m_{1}=1$, $m_{2}=1.25$, $k_{F}=10$ and all $\epsilon=1$
in the calculation.}

\end{figure}

For the OPS case with two screening layers, the interactions are respectively
$V_{1,2}=\frac{2\pi e^{2}}{\epsilon_{1,2}q}(1-e^{-qa})\sim2\pi e^{2}/\epsilon_{1,2}\left(a-a^{2}q/2\right)$
and $V_{x}=-\frac{2\pi e^{2}}{\epsilon_{x}q}e^{-qb}(1-e^{-qa})^{2}\sim-2\pi a^{2}e^{2}q/\epsilon_{x}$
for $q\ll a^{-1}$. Notice the removal of the $1/q$ singularity in
the interactions by OPS. We then obtain after solving the equation
the collective modes (up to order $q^{2}$) \begin{equation}
\omega_{1,2}=\sqrt{\frac{2\pi ae^{2}n_{1,2}}{m_{1,2}\epsilon_{1,2}}}\left(q-\frac{aq^{2}}{4}\right)\;.\end{equation}
 Notice that OPS effectively reduced the long-ranged Coulomb interaction
into short-ranged interactions resulting in two collective modes scaling
linearly with $q$. The collective modes represent separate collective
motion of the two layers because $\left(V^{x}\right)^{2}$ is of higher
order in $q$ than $V_{1}V_{2}$, and the inter-layer interaction
appears only to order $q^{3}$. For completeness, we have computed
numerically the collective modes spectrums at finite $q$ as shown
in Fig.\,\ref{fig:Plasmon-d-d}.

\begin{figure}
\includegraphics[width=1\columnwidth]{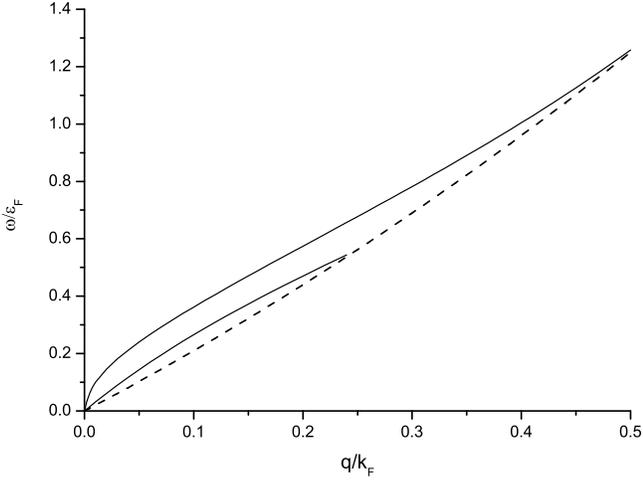}

\caption{\label{fig:singleplasmon}Plasmon excitations in EHBL with only one
screening layer. At small $q$, both plasmon modes scale linearly
with $q$ with a larger slope $\propto\sqrt{2b+a}$ for $\omega_{1}$.
We set $a=1$, $b=15$, $m_{1}=1$, $m_{2}=1.25$, $k_{F}=10$ and all
$\epsilon=1$ in the calculation.}

\end{figure}

With only one screening layer, the interactions are $V_{1}(q)\sim\frac{2\pi e^{2}}{\epsilon_{1}}\left((a+2b)-(a+2b)^{2}q/2\right)$,
$V_{2}(q)\sim\frac{2\pi e^{2}}{\epsilon_{2}}\left(a-a^{2}q/2\right)$
and $V_{x}(q)=-\frac{2\pi e^{2}}{\epsilon_{x}q}e^{-qb}(1-e^{-qa})\sim-2\pi ae^{2}/\epsilon_{x}$,
respectively at small $q$. The collective modes are given by (up
to order $q^{2}$) \begin{eqnarray}
\omega_{1} & = & \sqrt{\frac{2\pi(a+2b)e^{2}n_{1}}{m_{1}\epsilon_{1}}}\left(q-\frac{(a+2b)q^{2}}{4}\right)\nonumber \\
\omega_{2} & = & \sqrt{\frac{2\pi ae^{2}n_{2}}{m_{2}\epsilon_{2}}}\left(q-\frac{aq^{2}}{4}\right)\;.\end{eqnarray}
 Again there are two linear plasmon modes and effect of $V^{x}$ does
not enter until $q^{3}$. The main difference is that the electron-hole
layer separation $b$ enters the slope of $\omega_{1}$ mode ($\propto\sqrt{2b+a}$).
The numerically calculated plasmon spectrums are depicted in Fig.\,\ref{fig:singleplasmon}.

\section{Exciton Condensation and Wigner Crystalization}

In this section we study exciton condensation and Wigner crystalization
in an electron-hole liquid with OPS. The system without OPS has been
extensively studied for the search of exciton condensation. We shall
consider exciton condensation in a BCS type mean-field theory where
the exciton condensation is described by the order parameter $\left\langle c_{1k\uparrow}c_{2\bar{k}\downarrow}\right\rangle $
(1,2 are layer indices). For simplicity we assume the layers are doped
with equal amount of charges (with opposite signs) and the electrons
and holes are spin-polarized. Singlet pairing of excitons is implicitly
assumed.

The EHBL Hamiltonian in momentum representation is \begin{eqnarray}
H & = & \sum_{\alpha k}\xi_{k}^{\alpha}c_{\alpha k}^{\dagger}c_{\alpha k}+\sum_{pqk}V^{x}(k)c_{1p+k}^{\dagger}c_{2q-k}^{\dagger}c_{2q}c_{1p}\nonumber \\
 &  & +\frac{1}{2}\sum_{\alpha pqk}V^{\alpha}(k)c_{\alpha p+k}^{\dagger}c_{\alpha q-k}^{\dagger}c_{\alpha q}c_{\alpha p}\;,\end{eqnarray}
 where $\alpha=1,\!2$ is the layer index; $c_{k}$ ($c_{k}^{\dagger}$)
is the momentum $k$ fermion annihilation (creation) operator, $\xi_{k}^{\alpha}=\frac{k^{2}}{2m_{\alpha}}-\mu_{\alpha}$
is the electron or hole dispersion and $V^{\alpha}(k)$ ($V^{x}(k)$)
is the intralayer (interlayer) OPS effective interaction. Next we
employ the standard Hartree-Fock-Bogoliubov method \cite{HartreeFock}
to derive the mean field equations for exciton condensate. The Hartree-Fock
terms $\Sigma_{k}^{\alpha}=\sum_{q}V^{\alpha}(p-q)\left\langle c_{\alpha k}^{\dagger}c_{\alpha k}\right\rangle $
modify the particle dispersions $\xi_{k}^{\alpha}\rightarrow\xi_{k}^{\alpha}-\Sigma_{k}^{\alpha}$
and need to be solved self-consistently. Here we concentrate on the
effect of exciton binding on the Fermi surface and shall assume that
the self-energy can be captured by introducing effective masses $m_{\alpha}^{*}\left(\epsilon_{\alpha}\right)$
and renormalized chemical potentials $\mu_{\alpha}^{*}\left(\epsilon_{\alpha}\right)$,
i.e.\ $\xi_{k}^{\alpha}-\Sigma_{k}^{\alpha}\sim\frac{k^{2}}{2m_{\alpha}^{*}}-\mu_{\alpha}^{*}$.
With this approximation, we obtain the mean field Bogoliubov Hamiltonian
\begin{equation}
H_{\text{MF}}=\sum_{k\sigma}\left(\begin{array}{cc}
c_{1k}^{\dagger} & c_{2\bar{k}}\end{array}\right)\left(\begin{array}{cc}
\xi_{k}^{1} & -\Delta_{k}\\
-\Delta_{k} & -\xi_{k}^{2}\end{array}\right)\left(\begin{array}{c}
c_{1k}\\
c_{2\bar{k}}^{\dagger}\end{array}\right)\;,\end{equation}
 where \begin{equation}
\Delta_{k}=-\sum_{q}V^{x}(k-q)\left\langle c_{1k}c_{2\bar{k}}\right\rangle \label{eq:gapeq}\end{equation}
 is the exciton order parameter. $H_{\text{MF}}$ can be diagonalized
easily by the Bogoliubov transformation \begin{equation}
\left(\begin{array}{c}
c_{1k}\\
c_{2\bar{k}}^{\dagger}\end{array}\right)=\left(\begin{array}{cc}
u_{k} & \upsilon_{k}\\
-\upsilon_{k} & u_{k}\end{array}\right)\left(\begin{array}{c}
\gamma_{1k}\\
\gamma_{2\bar{k}}^{\dagger}\end{array}\right)\;,\end{equation}
 \begin{equation}
\left\{ \begin{array}{rcl}
u_{k}^{2} & = & \frac{1}{2}\left(1+\frac{\bar{\xi}_{k}}{E_{k}}\right)\;,\\
\upsilon_{k}^{2} & = & \frac{1}{2}\left(1-\frac{\bar{\xi}_{k}}{E_{k}}\right)\;,\end{array}\right.\end{equation}
 where $E_{k}=\sqrt{\left(\bar{\xi}_{k}\right)^{2}+\Delta_{k}^{2}}$,
$\bar{\xi}_{k}=\frac{1}{2}\left(\xi_{k}^{1}+\xi_{k}^{2}\right)\equiv\frac{k^{2}}{2m_{\text{eff}}}-\mu$,
where $m_{\text{eff}}^{-1}=(m_{1}^{*-1}+m_{2}^{*-1})/2$ and $\mu=(\mu_{1}^{*}+\mu_{2}^{*})/2$.
The ground state wavefunction is \begin{equation}
\left|\psi_{G}\right\rangle =\prod_{k}\left(u_{k}+\upsilon_{k}c_{1k}^{\dagger}c_{2\bar{k}}^{\dagger}\right)\left|0\right\rangle \;.\end{equation}
 where $\Delta_{k}$ is determined by the self-consistent equation
\begin{equation}
\Delta_{k}=-\frac{1}{2}\sum_{q}V^{x}(k-q)\frac{\Delta_{q}}{E_{q}}\;.\label{eq:self1}\end{equation}
 The equation is to be solved with the particle number constraint
\begin{equation}
n=\sum_{k}\upsilon_{k}^{2}\;,\label{eq:self2}\end{equation}
 where $\upsilon_{k}^{2}$ is the probability of finding an electron-hole
pair in state $k$ at the ground state. A zero-temperature phase diagram
can be determined by numerically solving eqs.\,\eqref{eq:self1}
and \eqref{eq:self2}.

To simplify calculation we assume further that exciton gap is momentum
independent $\Delta_{k}=\Delta$ and $\Delta$ is determined by minimizing
the ground state energy. We note that we are considering a band structure
with isotropic dispersion and the electron and hole Fermi surfaces
are perfectly nested. In this case, the exciton pairing gap $\Delta$
is always non-zero in the mean-field theory, although its value can
be very small. In reality the mean-field gap will be destroyed by
quantum fluctuations when it's magnitude is small \cite{exction-cond2},
but this is not reflected in a mean field theory. To capture this
physics qualitatively, we assume that the transition from the exciton
condensed state to the normal state occurs at $\Delta=10^{-5}\mu$.
Although quantitatively unreliable, this procedure allows us to examine
the effect of screening on the phase diagram semi-quantitatively as
we shall see below.

\begin{figure}
\includegraphics[width=1\columnwidth]{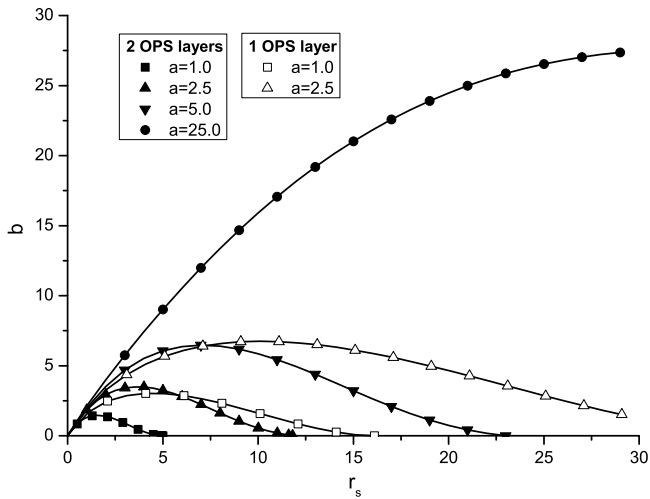}

\caption{\label{fig:d-vs-rs} The phase diagram for EHBL with two- and one-
screening layers for varied image charge separation $a$. $b$ is
the bilayer separation and $r_{s}\sim\frac{1}{k_{F}}$ is the dimensionless
average particle-particle separation in unit of the effective Bohr
radius $a_{B}=\epsilon_{x}\hbar^{2}/m_{\text{eff}}e^{2}$. Below the
transition lines $b_{c}(r_{s})$ an exciton gap $\Delta$ of magnitude
larger than $10^{-5}\mu$ is formed.}

\end{figure}

With the above criteria, the phase diagram for different average particle-particle
separation $r_{s}=\frac{1}{a_{B}}\sqrt{\frac{1}{\pi n}}$ ($n$ is
particle/hole density; $a_{B}=\epsilon_{x}\hbar^{2}/m_{\text{eff}}e^{2}$
is the effective Bohr radius of electron-hole pair) and transition
layer separation $b_{c}(r_{s})$ can be determined by solving the
self-consistent equations \eqref{eq:self1} and \eqref{eq:self2}.
We first consider EHBL with two screening layers. The result of calculation
is depicted in Fig.\,\ref{fig:d-vs-rs} for different separation
between the electron/hole and its image charge $a$ (filled symbols).

In the small $r_{s}$ (high density) regime, kinetic energy dominates
over potential energy and the exciton pairing gap goes to zero as
$r_{s}\rightarrow0$. In large $r_{s}$ or low density limit, the
exciton pairing is diminished due to the repulsive nature of the interlayer
OPS potential at short distance (see Fig.\,\ref{fig:EHBL-OPS}(f)).
This leads to a linear dependence of $V^{x}(k)$ versus $k$ at small
$k$ (see eq.\,\eqref{v3}). In this case, the gap equation eq.\,\eqref{eq:self1}
is of the form $\int_{0}^{k_{F}}\frac{k}{\sqrt{\xi_{k}^{2}+\Delta^{2}}}d^{2}k=\text{constant}$
for small gap $\Delta$, where $k_{F}\sim1/r_{s}$ and larger $r_{s}$
(smaller $k_{F}$) implies a smaller $\Delta$ to satisfy the equation.
The electrons and holes need to be placed closer to each other to
produce a large enough $\Delta$ and leads to the drop of $b_{c}(r_{s})$
at large $r_{s}$. As the screening separation $a$ increases, the
transition line shifts upward as the interlayer OPS potential is strengthened
which enhances pairing. For $a=25$ (comparable with $b$), the image
charge effect becomes negligible and potential becomes essentially
Coulomb-like which permits exciton formation for all $r_{s}$ we considered
(cf.\ Fig.\,1 in Ref.\,\cite{exction-cond2}). The main effect
of OPS potential is to suppress exciton pairing at low density.

Previous numerical study of the same EHBL with no screening layer
\cite{exction-cond2} reveals also an excitonic Wigner crystal
phase at large $r_{s}$. Wigner crystal is commonly formed in low
density (i.e.\ large $r_{s}$) electron liquid because of
domination of Coulomb repulsive potential energy ($\sim1/r_{s}$)
over kinetic energy ($\sim1/r_{s}^{2}$). To minimize the potential
energy the electron wavefunction {}``crystallizes'' to ensure
maximum separation between electrons which yields the Wigner
crystal phase. Here we argue that OPS suppresses the Wigner
crystal phase in two ways. Firstly, as shown above, exciton
formation is suppressed at large $r_{s}$ and thus the
\textit{excitonic} Wigner crystal is unlikely to form. On the
other hand, electronic Wigner crystals in separated layers are
also prohibited since introduction of OPS reduces the (intralayer)
potential energy and changes its scaling form to $\sim1/r_{s}^{3}$
(dipolar interaction, see eq.\,\eqref{eq:effintra2}) at large
particle separation $r\gg a$. In this case kinetic energy again
dominates at large $r_{s}$ and an usual electron/hole liquid phase
should occur. The situation is similar to the case as found in
Ref.\,\cite{Nonact-transport} where the electronic Wigner crystal
phase is destroyed by screening. We note, however that our simple
study cannot rule out the possibility of having a Wigner crystal
phase at some intermediate values of $r_{s}$ where the kinetic and
potential energies are of comparable magnitudes.

We now consider the situation of EHBL with only one screening layer
which may be easier to realize experimentally (Fig.\,\ref{fig:EHBL-OPS}(c)).
In this case we adopt eq.\,\eqref{v5} for interlayer interaction,
where $V^{x}(k)$ scales as constant at small $k$. We can again consider
the gap equation and argue similarly that the exciton phase boundary
would also drop at large $r_{s}$, as in the two-layer screening case.
Indeed we have solved the gap equations and find that the phase diagram
is qualitatively the same as the two OPS layer case except that the
area under the phase boundary $b_{c}(r_{s})$ is larger (see Fig.\,\ref{fig:d-vs-rs}
(open symbols)).

For the Wigner crystal phase, the ``asymmetric'' OPS introduces
some complications. First we note that an excitonic Wigner crystal
phase is also unlikely to occur at large $r_s$. However the system
may form a hybrid phase where a Wigner crystal is formed at layer
$1$ and electron/hole liquid phase remains for layer $2$ because screening
mainly affects layer $2$. To examine this possibility we check the
effective intralayer interaction after taking into account the screening
effect of the other charged layer (see eq.\,\eqref{eq:effintra2} in Appendix and
discussions thereafter). We see that the effective intralayer
interaction is mainly dominated by $V_{1,2}^{\text{intra}}(q)$,
and screening from the other layer is not important. Therefore, we
expect that at large $r_s$ kinetic energy again dominates and the
both layers are in the electron/hole liquid phase. Notice, however
that $V_{1}^{\text{intra}}(q)$ has a dipolar form only when
$r_s\sim r/a_B \gg b/a_B$ for layer 1. Thus for some large enough
$b/a_B$, a hybrid phase (Wigner crystal at layer $1$,
electron/hole liquid at layer $2$) may still occur at some
intermediate densities $b/a_B\gg r_s\gg 1$.

We see that OPS becomes important for low density electronic systems
due to change in scaling of the potential energy. Generally speaking,
for heterostructures, insulating behavior resulting from low carrier
density can be avoided by addition of metallic screening layers \cite{Nonact-transport}.
This method may be preferred over other methods like increasing carrier
density by dopants since dopants act like impurities and introduce
unnecessary scattering at low temperature.

\section{Strongly Correlated EHBL}

In strongly correlated materials, the basic electronic properties
are determined by the bandwidth, the on-site Coulomb interactions
$U$ and the charge transfer energy $E_{c}$. If such a ultra-thin
film, originally a Mott insulator, is placed close to a metal surface,
$U$ and $E_{c}$ can be strongly reduced by OPS \cite{thin_film_on_metal}.
When the bandwidth exceeds the suppressed $U$ and $E_{c}$, the insulating
film can undergo an insulator-metal phase transition. Furthermore,
if a heterostructure is formed, structural relaxation and local electronic
states may exist at the interfaces. For instance, in an interface
formed by $\mathrm{YBa_{2}Cu_{3}O_{7}}$ (YBCO) cuprate and metal
\cite{YBCO_metal_interface1,YBCO_metal_interface2}, the $\mathrm{CuO_{2}}$
plane near the interface (depletion layer) is intrinsically doped
by electronic reconstruction resulting in a strongly correlated electron
system with OPS interaction induced by the metal. We shall consider
here how OPS would affect the properties of this system.

The mean field analysis on effect of OPS can also be performed for
strongly correlated EHBL systems \cite{SLI-1,SLI-2} with a two-layer
\textit{t-J} type model. We assume here that the suppression of $U$ and $E_c$ induced by OPS
are not strong enough to destroy strong correlation, otherwise we
can simply apply the usual electron-hole liquid picture described
in previous section. Therefore the setting is similar to that shown
in Fig.\,\ref{fig:EHBL-OPS}(b) except that the electron-hole liquid
is replaced by a strongly correlated EHBL with holons and doublons
and the excitons are formed by holon-doublon pairs instead of electron-hole
pairs. A mean field calculation similar to that of Ref.\,\cite{SLI-1}
can be carried out by applying the slave-boson mean field theory to
the two-layer \textit{t-J} model. The main difference is that the
on-site interlayer interaction $V_{0}\sum_{i}b_{1i}^{\dagger}b_{1i}b_{2i}^{\dagger}b_{2i}$
is replaced by the OPS effective interaction $\sum_{ij}V_{ij}^{x}b_{1i}^{\dagger}b_{1i}b_{2j}^{\dagger}b_{2j}$,
where $b_{\alpha i}$ ($b_{\alpha i}^{\dagger}$) is the bosonic holon
$(\alpha=1)$ or doublon $(\alpha=2)$ annihilation (creation) operator
of layer $\alpha$ at site $i$. The OPS interaction is then decoupled
as \begin{eqnarray*}
\sum_{ij}V_{ij}^{x}b_{1i}^{\dagger}b_{1i}b_{2j}^{\dagger}b_{2j} & \mathcal{\overset{F}{\longrightarrow}} & \sum_{pqk}V_{k}^{x}b_{1p}^{\dagger}b_{2q}^{\dagger}b_{2q+k}b_{1p-k}\\
 & \approx & \sum_{p}\Delta_{p}^{b}\left(b_{1p}b_{2\bar{p}}+b_{2\bar{p}}^{\dagger}b_{1p}^{\dagger}\right)\\
 &  & -\sum_{p}\Delta_{p}^{b}\left\langle b_{1p}b_{2\bar{p}}\right\rangle \;,\end{eqnarray*}
 where $\Delta_{p}^{b}=\sum_{q}V_{p-q}^{x}\left\langle b_{1q}b_{2\bar{q}}\right\rangle $
is the exciton pairing. Assuming that $\Delta_{p}^{b}=\Delta^{b}\delta(p)$
is homogeneous in space, we obtain a mean field Hamiltonian which
is of the same form as in previous study \cite{SLI-1} for on-site
interaction $V_{0}$ with $\left\langle b_{1i}b_{2i}\right\rangle \sim\sum_{k}\left\langle b_{1k}b_{2\bar{k}}\right\rangle $.
Since in terms of exciton pairing the attract-repel behavior renders
the interlayer OPS interaction resembling an on-site interaction (see
Fig.\,\ref{fig:EHBL-OPS}(f)), we expect that the mean field phase
diagrams in both case are qualitatively the same. The introduction
of OPS interaction solely shifts the exciton phase boundary due to
a reduction of interaction strength, as in the case of usual electron-hole
liquid.

We next comment on the possibility of forming spatially inhomogeneous
phases. One example of such inhomogeneity is charge corrugation in
the form of stripes. By applying mean field theory to \textit{t-J}
model with long range Coulomb interaction $V_{c}\sum_{i\neq j}\frac{1}{r_{ij}}n_{i}n_{j}$,
it is shown that stripes are preferred to minimize the exchange $J$
term \cite{LeeDH-stripes}. In particular, it is the decoupling of
the exchange term into the anti-ferromagnetic channel $m_{i}$ that
drives the stripe formation, while the Coulomb interaction controls
the spacing evolution of stripes with doping. Moreover, the stripes
spacing increases as the doping $\delta$ decreases. The effect of
OPS on stripes is two-fold. Firstly, OPS weakens the on-site repulsion
$U$ \cite{Inc-J} and thus the superexchange $J\sim t^{2}/U$ term
is enhanced (assuming that strong correlation is still intact). Consequently,
the stripes phase is strengthened. On the other hand, Coulomb interaction
tends to smooth out the charge density, while a dipolar interaction
($V\sim1/r^{3}$ for large $r$) would be less effective and a more
inhomogeneous phase would be preferred. Notice that extreme charge
inhomogeneity like phase separation \cite{phase_sep} is not likely
since the OPS interaction scales like $1/r$ for small $r$ and still
suppresses phase separation.

\section{Summary}

We have constructed a dipolar interaction for OPS effect of metallic
layer in heterostructures and have justified the construction by a
RPA calculation. The OPS interaction is expected to be present rather
generally at interfaces with metallic layers. We apply the OPS interaction
to EHBL system and find that OPS mainly affects the electronic properties
in the low density regime. Our conclusion is not restricted to EHBL
since the behavior is mainly due to the modification of the interaction
scaling from $1/r$ (Coulomb) to $1/r^{3}$ (dipole) at distance of
large $r$. OPS might be employed to eliminate Wigner-crystal like
behavior at low temperatures. For strongly correlated electron systems,
OPS mainly affects the magnetic channel by reducing the Hubbard $U$
and charge transfer energy $E_{c}$. The reduction of $U$ may drive
the system into usual electron liquid. Furthermore, the reduction
in interaction range may drive the system into an inhomogeneous state.
\begin{acknowledgments}
We acknowledge Prof.\ P. A. Lee and Prof.\ N. Nagaosa for insightful
comments and Dr.\ Y. Zhou, Dr.\ X. Y. Feng, C. K. Chan and Z. X.
Liu for helpful discussions.
\end{acknowledgments}

\appendix*

\section{Justification of OPS interactions by RPA}

In this appendix, we employ RPA to justify the image-charge picture
of OPS interactions. The RPA method enables us to obtain an effective
interaction by {}``integrating out'' the screening layers.

\begin{figure}
\includegraphics[width=0.8\columnwidth]{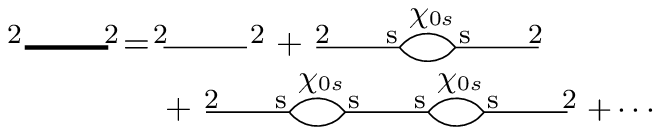}

\caption{\label{fig:Feyn}Diagram for construction of OPS interactions. Thin
lines are bare interactions, bubble is the density-density response
function $\chi_{0s}$ and the thick line is the resulting effective
interaction.}

\end{figure}

First we consider a charged layer $2$ with a metallic screening layer
$s$ separated from layer $2$ by distance $a/2$, as shown in Fig.\,\ref{fig:EHBL-OPS}(c).
We can write down the effective intralayer interaction of layer $2$
after taking into account the effect of screening by the metallic
layer (see Fig.\,\ref{fig:Feyn})

\begin{eqnarray*}
V^{\text{intra}}(q) & = & V_{2}(q)+\frac{V_{2s}(q)\chi_{0s}V_{s2}(q)}{1-\chi_{0s}V_{ss}(q)}\\
 & = & 2\pi e^{2}\left(\frac{1}{q}-\frac{e^{-aq}}{q}\frac{2\pi e^{2}N_{F}}{q+2\pi e^{2}N_{F}}\right)\\
 & \approx & \frac{2\pi e^{2}}{q}\left(1-e^{-aq}\right)\;,\end{eqnarray*}
where $V_{2}(q)=V_{ss}(q)=2\pi e^{2}/q$ are the bare Coulomb interactions
of layer $2$ and screening layer $s$, $V_{2s}(q)=V_{s2}(q)=2\pi e^{2}e^{-aq/2}/q$
is the interlayer Coulomb interaction between the layers, and $\chi_{0s}=\chi_{0s}(q\rightarrow0,\omega=0)=-N_{F}$
is the $q\rightarrow0$ static density-density response function \cite{G&V}
of layer $s$, $N_{F}$ is density of states at the Fermi surface.
Notice we have assumed that $q$ is small (long wavelength limit)
in writing down the interactions and therefore \[
F_{0}\equiv\frac{2\pi e^{2}N_{F}}{q+2\pi e^{2}N_{F}}\approx1\;.\]
 This approximation is valid if the charge density $n_{2}$ of layer
2 is much less than the density of the screening layer $n_{s}$ and
$q\lesssim\sqrt{n_{2}}\ll\sqrt{n_{s}}$. We shall take the same limit
in the following derivations. This gives eq.\,\eqref{v2}.

Similarly we can construct the interlayer interaction eq.\,\eqref{v5}:

\begin{eqnarray*}
V^{x,1}(q) & = & V_{12}(q)+\frac{V_{1s}(q)\chi_{0s}V_{s2}(q)}{1-\chi_{0s}V_{ss}(q)}\\
 & = & 2\pi e^{2}\left(\frac{e^{-bq}}{q}-\frac{e^{-(a+b)q}}{q}F_{0}\right)\\
 & \approx & 2\pi e^{2}\frac{e^{-bq}}{q}\left(1-e^{-aq}\right)\;,\end{eqnarray*}
 where $V_{12}(q)=2\pi e^{2}e^{-bq}/q$, $V_{1s}(q)=2\pi e^{2}e^{-(a/2+b)q}/q$
and $V_{s2}(q)=2\pi e^{2}e^{-aq/2}/q$ are the bare interlayer interactions
between the pair of layers ($1$,$2$), $(1,s)$ and $(s,2)$ respectively.
For two screening layers (Fig.\,\ref{fig:EHBL-OPS}(b)), we assume
that layer $2$ and $s$ form an effective system $2^{\prime}$ and
thus we can adopt $V^{x,1}$ as the {}``bare'' interlayer interactions
in the following:

\begin{eqnarray*}
V^{x,2}(q) & = & V^{x,1}(q)+\frac{V_{1s}(q)\chi_{0s}V_{s2^{\prime}}(q)}{1-\chi_{0s}V_{ss}(q)}\\
 & = & 2\pi e^{2}\left(1-e^{-aq}\right)\left(\frac{e^{-bq}}{q}-\frac{e^{-(a+b)q}}{q}F_{0}\right)\\
 & \approx & 2\pi e^{2}\frac{e^{-bq}}{q}\left(1-e^{-aq}\right)^{2}\;,\end{eqnarray*}
 where $V_{s2^{\prime}}(q)=V^{x,1}(b\rightarrow b+\frac{a}{2})=\frac{1}{q}e^{-(\frac{a}{2}+b)q}\left(1-e^{-aq}\right)$.
The validity of assuming the effective system $2^{\prime}$ is based
on the choice of $b\gg a$. This gives eq.\,\eqref{v3}.

Here we derive the effective intralayer interaction $V_{1,\text{eff}}^{\text{intra}}(q)$
with two-layer OPS (Fig.\,\ref{fig:EHBL-OPS}(b)) taking into account
the screening of system $2^{\prime}$ (i.e.\ integrated out all the
screening by $2$, $s1$ and $s2$):

\begin{equation}
V_{1,\text{eff}}^{\text{intra}}(q)=V_{1}^{\text{intra}}(q)+\frac{\chi_{02}\left(V^{x,2}(q)\right)^{2}}{1-V_{2}^{\text{intra}}(q)\chi_{02}}\;.\label{eq:effintra2}\end{equation}
In the small $q$ limit, $V^{x,2}$ and $V_{2}^{\text{intra}}$ scale
as $q$ and constant respectively. The second term due to screening
is of higher order in $q$ and thus it cannot alter the scaling of
the $V_{1}^{\text{intra}}(q)$ term ($\sim\text{constant}$). One can
repeat the analysis for the one-layer OPS case (see
Fig.\,\ref{fig:EHBL-OPS}(c)) and the scaling of the effective intralayer
interaction $V_{1,2}^{\text{intra}}(q)$ in the lowest order of $q$ is
not affected by screening of the opposite charged layer.


\begin{thebibliography}{17}
\bibitem{AMillis}S. Okamoto and A. J. Millis, Nature \textbf{428},
630 (2004).

\bibitem{Xconfine92}X. Hong, T. Ishihara and A. V. Nurmikko, Phys.
Rev. B \textbf{45}, 6961 (1992).

\bibitem{Xconfine95}E. A. Mulijarov, S. G. Tikhodeev and N. A. Gippius,
Phys. Rev. B \textbf{51}, 14370 (1995).

\bibitem{Nonact-transport}J. Huang, D. S. Novikov, D. C. Tsui, L.
N. Pfeiffer and K. W. West, Phys. Rev. B \textbf{74}, 201302(R) (2006).
See also, L. H. Ho \textit{et al.}, Phys. Rev. B 77, 201402(R) (2008).

\bibitem{HoLH}L. H. Ho, A. P. Micolich, A. R. Hamilton and O. P.
Sushkov, Phys. Rev. B \textbf{80}, 155412 (2009).

\bibitem{exciton-cond}Sen Yang, A. T. Hammack, M. M. Fogler and L.
V. Butov, Phys. Rev. Lett. \textbf{97}, 187402 (2006).

\bibitem{thin_film_on_metal} S. Altieri, L. H. Tjeng and G. A. Sawatzky,
Thin Solid Films \textbf{400}, 9-15 (2001).

\bibitem{YBCO_metal_interface1} U. Schwingenschlögl and C. Schuster,
Europhys. Lett. \textbf{77}, 37007 (2007).

\bibitem{YBCO_metal_interface2} U. Schwingenschlögl and C. Schuster,
Phys. Rev. B \textbf{79}, 092505 (2009).

\bibitem{G&V}See, e.g., G. F. Giuliani and G. Vignale, \textit{Quantum
Theory of the Electron Liquid} (Cambridge University Press, Cambridge,
2005).

\bibitem{HartreeFock}See, e.g., Xuejun Zhu, P. B. Littlewood, Mark
S. Hybertsen and T. M. Rice, Phys. Rev. Lett. \textbf{74}, 1633 (1995).

\bibitem{exction-cond2}S. De Palo, F. Rapisarda and Gaetano Senatore,
Phys. Rev. Lett. \textbf{88}, 206401 (2002).

\bibitem{SLI-1} Jung Hoon Han and Chenglong Jia, Phys. Rev. B \textbf{74},
075105 (2006).

\bibitem{SLI-2} T. C. Ribeiro, A. Seidel, J. H. Han and D.-H. Lee,
Europhys. Lett. \textbf{76}, 891 (2006).

\bibitem{LeeDH-stripes}Junghoon Han, Qiang-Hua Wang and Dung-Hai
Lee, Int. J. of Mod. Phys. B \textbf{15}, 1117 (2001).

\bibitem{Inc-J}S. Altieri \textit{et al.}, Phys. Rev. B \textbf{79},
174431 (2009).

\bibitem{phase_sep}V. J. Emery, S. A. Kivelson and H. Q. Lin, Phys.
Rev. Lett. \textbf{64}, 475 (1990).
\end{thebibliography}
\end{document}